\documentclass[fleqn,usenatbib]{mnras}
\usepackage{newtxtext,newtxmath}
\usepackage[T1]{fontenc}
\DeclareRobustCommand{\VAN}[3]{#2}
\let\VANthebibliography\thebibliography
\def\thebibliography{\DeclareRobustCommand{\VAN}[3]{##3}\VANthebibliography}

\usepackage{graphicx}	% Including figure files
\usepackage{amsmath}	% Advanced maths commands
\usepackage{tablefootnote}
\usepackage{xcolor}

\title[Locating the absorbing gas in BCGs]{Does absorption against AGN reveal supermassive black hole accretion?}

\author[Tom Rose et al.]{
Tom Rose$^{1,2}$\thanks{E-mail: thomas.rose@uwaterloo.ca},
B. R. McNamara$^{1,2}$,
F. Combes$^{3}$,
A. C. Edge$^{4}$,
A. C. Fabian$^{5}$,
M. Gaspari$^{6}$,\newauthor
H. Russell$^{7}$,
P. Salom\'e$^{3}$,
G. Tremblay$^{8}$,
G. Ferland$^{9}$
\\
$^{1}$Department of Physics and Astronomy, University of Waterloo, Waterloo, ON N2L 3G1, Canada \\
$^{2}$Waterloo Centre for Astrophysics, Waterloo, ON N2L 3G1, Canada \\
$^{3}$LERMA, Observatoire de Paris, PSL Research Univ., College de France, CNRS, Sorbonne Univ., Paris, France\\
$^{4}$Centre for Extragalactic Astronomy, Durham University, DH1 3LE, UK\\
$^{5}$Institute of Astronomy, Cambridge University, Madingly Rd., Cambridge, CB3 0HA, UK\\
$^{6}$Department of Astrophysical Sciences, Princeton University, Princeton, NJ 08544, USA\\
$^{7}$School of Physics \& Astronomy, University of Nottingham, Nottingham, NG7 2RD, UK\\
$^{8}$Harvard-Smithsonian Center for Astrophysics, 60 Garden St., Cambridge, MA 02138, USA\\
$^{9}$Department of Physics and Astronomy, University of Kentucky, Lexington, Kentucky 40506-0055, USA\\
}

\date{Accepted XXX. Received YYY; in original form ZZZ}

\pubyear{2022}

\begin{document}
\label{firstpage}
\pagerange{\pageref{firstpage}--\pageref{lastpage}}
\maketitle

% Abstract of the paper
\begin{abstract}
Galaxies often contain large reservoirs of molecular gas which shape their evolution. This can be through cooling of the gas -- which leads to star formation, or accretion onto the central supermassive black hole -- which fuels AGN activity and produces powerful feedback. Molecular gas has been detected in early-type galaxies on scales of just a few tens to hundreds of solar masses by searching for absorption against their compact radio cores. Using this technique, ALMA has found absorption in several brightest cluster galaxies, some of which show molecular gas moving towards their galaxy's core at hundreds of km/s. In this paper we constrain the location of this absorbing gas by comparing each galaxy's molecular emission and absorption. In four galaxies, the absorption properties are consistent with chance alignments between the continuum and a fraction of the molecular clouds visible in emission. In four others, the properties of the absorption are inconsistent with this scenario. In these systems the absorption is likely produced by a separate population of molecular clouds in close proximity to the galaxy core and with high inward velocities and velocity dispersions. We thus deduce the existence of two types of absorber, caused by chance alignments between the radio core and: (i) a fraction of the molecular clouds visible in emission, and (ii) molecular clouds close to the AGN, in the process of accretion. We also present the first ALMA observations of molecular emission in S555, Abell 2390, RXC J1350.3+0940 and RXC J1603.6+1553 -- with the latter three having $\textnormal{M}_{\textnormal{mol}} >10^{10}\textnormal{M}_{\odot}$.
\end{abstract}

% Select between one and six entries from the list of approved keywords.
% Don't make up new ones.
\begin{keywords}
galaxies: ISM -- Galaxies, (galaxies:) quasars: absorption lines -- Galaxies, ISM: molecules -- Interstellar Medium (ISM), Nebulae
\end{keywords}

%%%%%%%%%%%%%%%%%%%%%%%%%%%%%%%%%%%%%%%%%%%%%%%%%%

%%%%%%%%%%%%%%%%% BODY OF PAPER %%%%%%%%%%%%%%%%%%

\section{Introduction}
Most early-type galaxies at the centres of clusters lie within a hot, X-ray emitting atmosphere. The molecular gas which condenses as this atmosphere cools may contribute to star formation, or migrate towards the centre of the galaxy where it can fuel AGN activity. As a result of this fuelling, the AGN produces powerful radio jets and lobes which propagate energy into the galaxy cluster. This can then prevent further gas cooling and stifle the formation of molecular gas \citep[][]{McNamara2016, Morganti2017, Gaspari2020}. In massive galaxies, the molecular gas therefore acts as a medium through which many different processes are linked, so it is integral to our understanding of the galaxies and clusters. 

The properties of this cooled gas have been predicted from a theoretical perspective \citep[e.g.][]{Nulsen2005,Pizzolato2005, McNamara2016}, studied with simulations \citep[e.g.][]{Gaspari2013}, and directly observed with emission lines \citep[e.g.][]{Crawford1999, Jaffe2005, Edge2002, Donahue2011, Olivares2019}. These wide-ranging studies cover regions as large as clusters, and as small as individual clouds. 

Theories and simulations typically cover a wide range of spatial scales, but observational studies tend to be more limited because molecular emission lines are relatively weak. This means they can only detect gas in large quantities. Outside our own galaxy, they therefore struggle to reveal how it behaves in compact regions. To some degree, this issue has been tackled with studies of molecular absorption lines. These are the shadows of gas clouds which lie in front of a bright and compact background radio source, such as a quasar. Observing molecular gas with this technique was pioneered with studies of the chance alignment of a foreground galaxy and the distant radio loud quasar PKS1830-211 \citep[][]{Wiklind1996a}. Several similar absorption line systems have been found, but due to the necessity of a chance alignment between a galaxy and a bright background continuum source, the number of these detections remains small \citep[other examples include][]{Muller2008, Muller2013}.

\begin{table*}
    \caption{The ALMA observations presented in this paper.}
    \centering
    \begin{tabular}{lccccccccr}
    \hline
    Source & Ra & Dec & Ang. res. & FOV & Int. Time & Obs. date & PWV & Line & Project code \\
     & (J2000) & (J2000) & (") & (") & (s) & (yyyy-mm-dd) & (mm) & & \\
    \hline
    NGC6868 & 20:09:54.1 & -48:22:46.3 & 0.111 & 25 & 1421 & 2021-11-20 & 1.32 & CO(2-1) & 2021.1.00766.S\\
    S555 & 05:57:12.5 & -37:28:36.5 & 0.87 & 56 & 2812 & 2018-01-23 & 2.27 & CO(1-0) & 2017.1.00629.S \\
    Hydra-A & 09:18:05.7 & -12:05:44.0 & 2.07 & 57 & 2661 & 2018-07-18 & 2.78 & CO(1-0) & 2017.1.00629.S \\
    Hydra-A & 09:18:05.6 & -12:05:44.0 & 0.262 & 26 & 5745 & 2018-10-30 & 0.96 & CO(2-1) & 2018.1.01471.S \\
    A2390 & 21:53:36.8 & +17:41:43.7 & 0.52 & 59 & 8014 & 2018-01-07 & 2.08 & CO(1-0) & 2017.1.00629.S \\
    A1644 & 12:57:11.6 & -17:24:34.1 & 2.26 & 57 & 2752 & 2018-08-21 & 1.43 & CO(1-0) & 2017.1.00629.S \\
    IC 4296 & 13:36:39.1 & -33:57:57.3 & 0.40 & 25 & 1724 & 2014-07-23 & 1.64 & CO(2-1) & 2013.1.00229.S \\
    Abell 2597 & 23:25:19.7 & -12:07:27.7 & 0.43 & 26 & 10886 & 2013-11-17 & 1.59 & CO(2-1) & 2012.1.00988.S \\
    NGC5044 & 13:15:24.0 & -16:23:07.6 & 0.74 & 54 & 2388 & 2018-09-20 & 0.49 & CO(1-0) & 2017.1.00629.S \\
    NGC5044 & 13:15:24.0 & -16:23:07.6 & 1.89 & 25 & 1421 & 2012-01-13 & 1.44 & CO(2-1) & 2011.0.00735.S \\
    RXC J1350.3+0940 & 13:50:22.1 & +09:40:10.7 & 0.89 & 61 & 5625 & 2018-09-16 & 0.66 & CO(1-0) & 2017.1.00629.S \\
    MACS 1931.8-2634 & 19:31:49.6 & -26:34:33.0 & 0.34 & 63 & 5322 & 2018-01-02 & 3.12 & CO(1-0) & 2017.1.00629.S \\
    RXC J1603.6+1553 & 16:03:38.1 & +15:54:02.4 & 0.83 & 60 & 1452 & 2018-09-16 & 0.82 & CO(1-0) & 2017.1.00629.S \\
    \hline
    \end{tabular}
    \label{tab:observations}
\end{table*}

\begin{table*}
    \caption{Properties of the ALMA images presented in this paper, produced using the observations listed above.}
    \centering
    \begin{tabular}{lccccc}
    \hline
    Source & Continuum Flux Density & Continuum Frequency & RMS Noise & Channel Width & Beam Size \\
     & (mJy) & (GHz) & (mJy/beam) & (km/s) & (")\\
    \hline
    NGC6868 & 38 & 107 & 0.9 & 2.6 & $0.17 \times 0.14$ \\
    S555 & 31 & 103 & 0.5 & 2.6 & $0.64 \times 0.53$ \\
    Hydra-A & 99 & 103 & 1.0 & 2.7 & $2.32 \times 1.65$ \\
    Hydra-A & 70 & 218 & 0.9 & 0.7 & $0.27 \times 0.20$ \\
    A2390 & 21 & 98 & 0.3 & 3.1 & $0.71 \times 0.57$ \\
    A1644 & 63 & 103 & 0.7 & 2.6 & $2.27 \times 1.54$ \\
    IC 4296 & 191 & 234 & 0.5 & 1.3 & $0.63 \times 0.57$ \\
    Abell 2597 & 14.6 & 221 & 0.9 & 0.7 & $0.74 \times 0.55$ \\
    NGC5044 & 31 & 107 & 0.6 & 2.5 & $0.54 \times 0.47$ \\
    NGC5044 & 50 & 228 & 1.3 & 0.6 & $2.15 \times 1.24$ \\
    RXC J1350.3+0940 & 25 & 95 & 0.3 & 2.9 & $1.17 \times 0.84$ \\
    MACS 1931.8-2634 & 8.0 & 92 & 0.2 & 3.4 & $1.24 \times 0.98$ \\
    RXC J1603.6+1553 & 108 & 97 & 0.6 & 2.8 & $1.03 \times 0.79$ \\
    \hline
    \end{tabular}
    \label{tab:images}
\end{table*}

More recently, several systems have been found which use a galaxy's own radio core as a backlight, something for which only ALMA has been able provide the necessary angular resolution and sensitivity \citep[][]{David2014, Tremblay2016, Rose2019a, Rose2019b, Rose2020}. In these studies, the compactness of the background radio source is particularly important because it makes it possible to detect gas on small scales. For example, in the brightest cluster galaxy Hydra-A, individual clouds of just a few tens of solar masses and with velocity dispersions of $\sim$ 1 - 5 km/s \citep[][]{Rose2020} have been found. In this respect, and in terms of their column densities, the individual clouds detected in this massive galaxy have remarkably similar properties to those seen in the Milky Way. This is surprising considering the extreme differences between the two galaxies in terms of their mass, star formation rates, AGN activity and jet powers.

Absorption lines detected using a galaxy's own radio core as a backlight are particularly advantageous because the redshift of the line can be used to infer inward or outward movement of the gas. \citet{Rose2019b} found that from the 9 known absorption line systems of this type, with a total of 15 individually resolved absorption regions detected, the gas tends to be in motion towards the galaxy core. Additionally, it has been suggested that the absorbing gas is likely in the innermost few hundred parsecs of the galaxies \citep[][]{Rose2019b}. In cases where the gas has inward velocities of hundreds of km/s, the suggested proximity to the galaxy cores has been used to infer ongoing supermassive black hole accretion \citep[e.g.][]{Tremblay2016, Rose2019b}.

In this paper we constrain the likely location of the gas detected via molecular absorption lines  in eight brightest cluster galaxies. Seven of these systems were part of the \citet[][]{Rose2019b} ALMA survey which aimed to detect CO(1-0) absorption against each galaxy's central continuum source. This survey's targets were chosen because they had the strongest X-ray emission among the brightest cluster galaxies visible to ALMA. This meant bright continuum sources could be observed without orientation effects biasing the potential detections of absorption lines. In two of those sources, NGC 5044 and Abell 2597, no CO(1-0) absorption was detected. However, CO(2-1) absorption against their radio cores had previously been found \citep[][]{David2014, Tremblay2016}. The eighth galaxy presented in this paper, IC 4296, was identified as an absorber when its CO(2-1) emission was observed by ALMA \citep[][]{Ruffa2019}.

We constrain the location of the detected gas by analysing galaxies in which both molecular absorption and emission features have been detected. In the majority of cases, the absorption lines we analyse have previously been published, and some have also had their molecular emission lines studied in detail. However, we now use the data to investigate links between the molecular gas visible in absorption and the molecular gas visible in emission. We also present CO emission data for three galaxies which were part of the \citet{Rose2019b} survey hunting for molecular absorption lines, but in which only emission was found. These galaxies have molecular masses of $>10^{10}\textnormal{M}_{\odot}$, but lack absorption lines.

Throughout this paper we assume $\Lambda$ cold dark matter cosmology with $\Omega_{\textnormal{M}} =0.3$ and $\Omega_{\Lambda} = 0.7$, and a Hubble constant of H$_{0}$ = 70 km/s/Mpc.

\section{Observations and data reduction}

We present ALMA observations of molecular gas in 11 brightest cluster galaxies at redshifts in the range z = 0.009 - 0.35. Details of these observations are shown in Table \ref{tab:observations}. All but one have previously appeared in multiple publications, the most relevant of which are as follows. The molecular emission and absorption lines of NGC5044 have been studied in several papers, including \citet{David2014}, \citet{Temi2018} and \citet{Schellenberger2020}. Similar work focusing on Abell 2597 has also been carried out by \citet{Tremblay2018}. The molecular absorption lines of S555, Abell 2390, Abell 1644 and NGC6868 have been presented in \citet{Rose2019b}, but little attention was paid to their emission lines. This paper also presented the CO(1-0) emission lines of RXC J1350.3+0940, MACS 1931.8-2634 and RXC J1603.6+1553, but they were not studied in detail. Recently, \citet{Baek2022} carried out a multi-wavelength study of Abell 1644, which includes the study of its molecular emission and absorption lines. Hydra-A has the largest number of molecular lines detected in any of our sources, with absorption from 11 different molecular species having been found by \citet{Rose2020}. The molecular emission of Hydra-A has also been presented by \citet{Rose2019a}. IC 4296 has absorption and emission lines which have been presented by \citet{Ruffa2019}.

All observations were handled using \texttt{CASA} version 6.4.0.16, a software package which is produced and maintained by the National Radio Astronomy Observatory (NRAO) \citep{CASA}. Self-calibrated measurement sets of the data were provided upon request to The European ALMA Regional Centre. From these, we produced continuum subtracted images using the CASA tasks \texttt{tclean} and \texttt{uvcontsub}. In \texttt{tclean}, we have used a hogbom deconvolver, natural weighting and produced images with the smallest possible channel width. In Table \ref{tab:images} we show they key properties of our reduced images, including continuum flux densities estimated using channels free of emission and absorption.

When converting the frequencies of the CO spectra to velocities, we use rest frequencies of \mbox{$f_{\textnormal{CO(1-0)}} = 115.271208$ GHz} and \mbox{$f_{\textnormal{CO(2-1)}} = 230.538000$ GHz}.

\begin{figure*}
	\includegraphics[width=\textwidth]{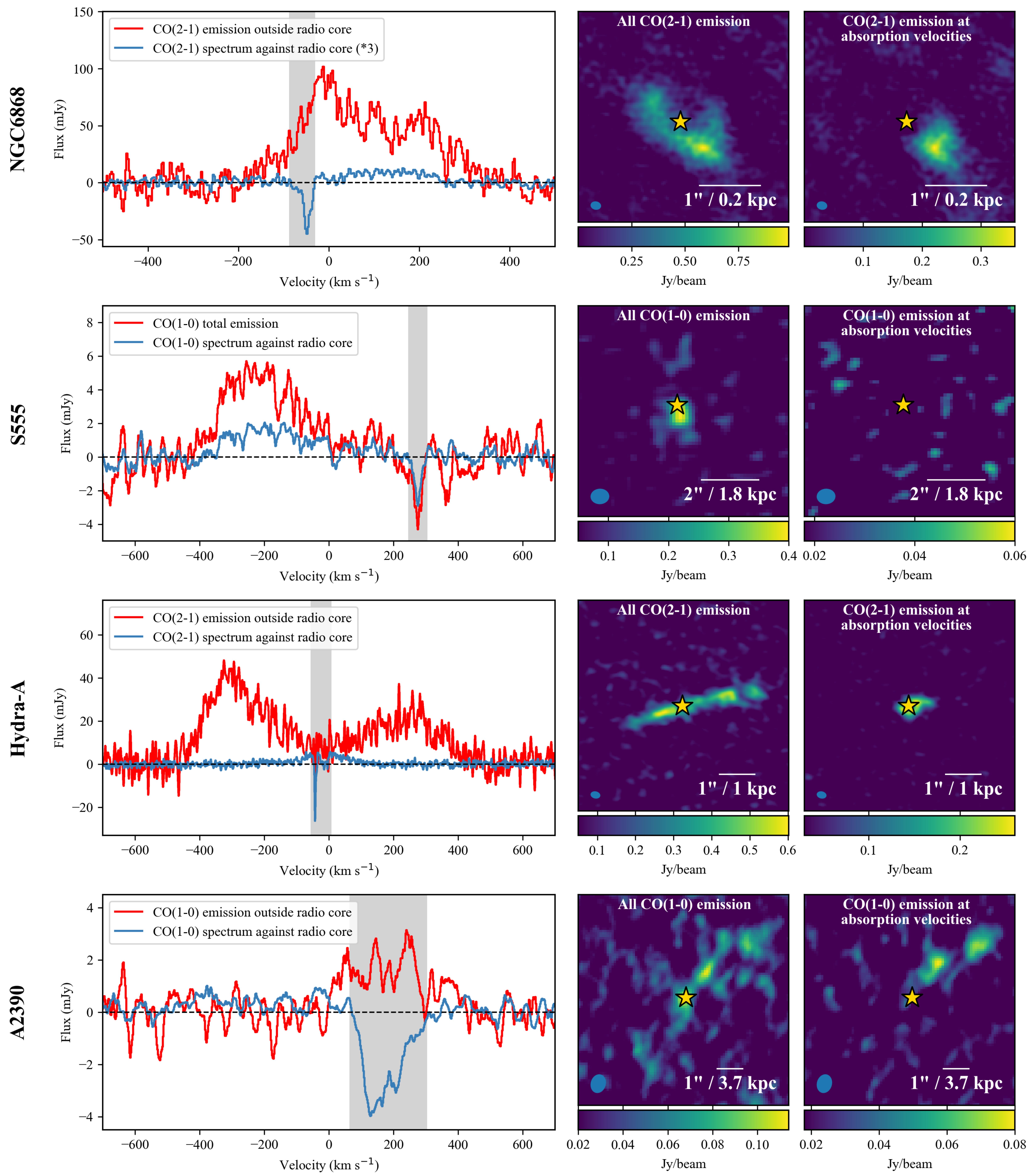}
    \caption{Galaxies with molecular absorption lines seen against their bright and compact radio core. \textbf{Left}: Red lines show CO spectra extracted from the region over which molecular gas is visible. This sometimes excludes the core region when the absorption is particularly strong (these cases are made clear in the legends). Blue spectra, which show the molecular absorption, are extracted from an approximately beam sized region centred on each object's continuum source (the location of which is indicated by the yellow star). \textbf{Right}: CO intensity maps made using (i) the full range of velocity channels in which  molecular emission is observed, and (ii) the velocity channels in which molecular absorption is observed, as indicated by the grey band.}
    \label{fig:main_fig_1}
\end{figure*}

\begin{figure*}
	\includegraphics[width=\textwidth]{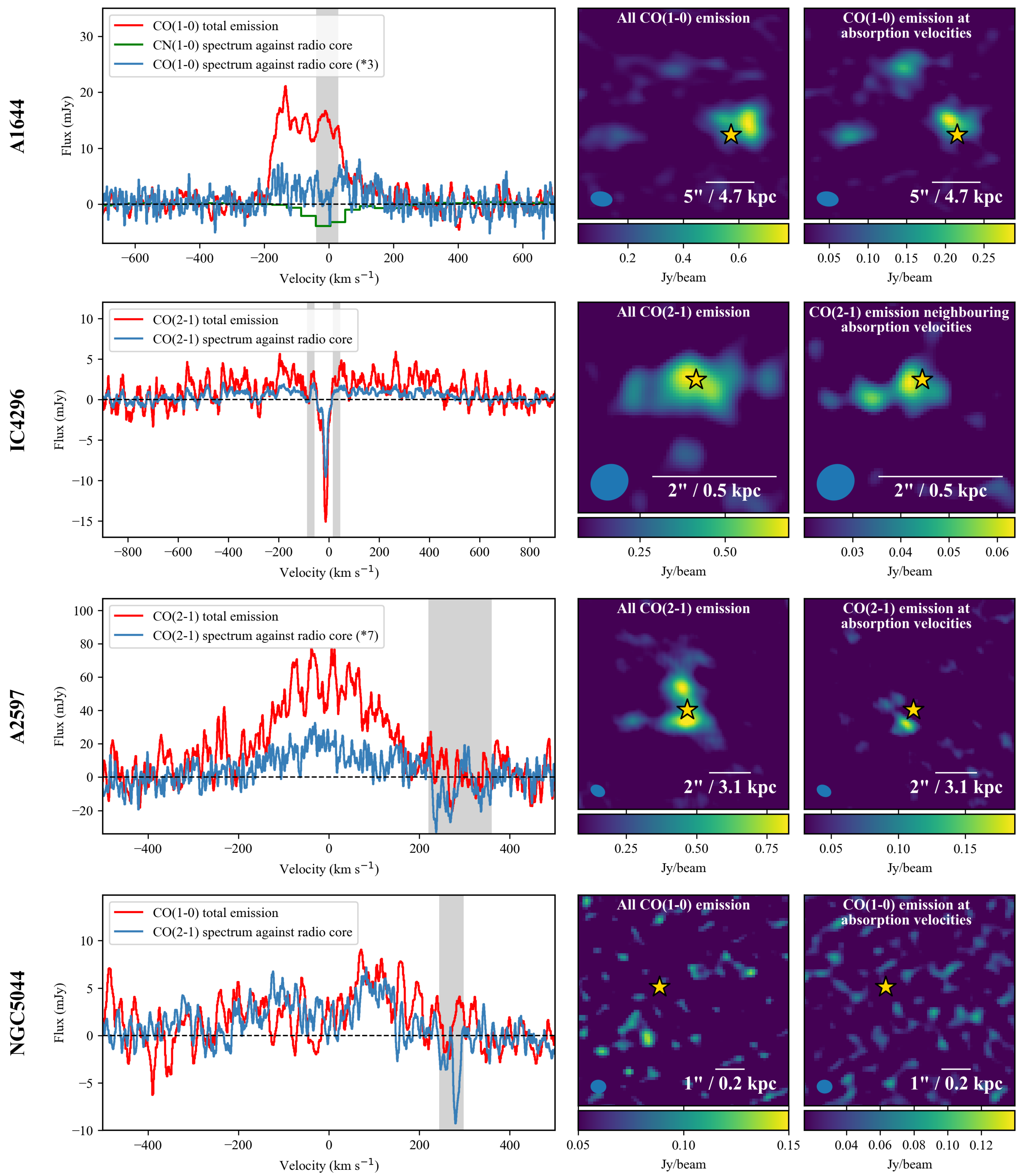}
    \caption{Continued from Fig. \ref{fig:main_fig_1}. Additional information: \textbf{A1644} has a complex CO(1-0) profile against its radio core which contains both emission and absorption. We therefore include the much stronger CN absorption line as a visual guide. However, this has much lower spectral resolution and contains hyperfine structure, which increases its apparent width. The intensity map on the right of \textbf{IC4296} is made with velocity channels neighbouring the absorption line. Using only the velocity channels in which absorption is seen is not possible because the emission is only marginally extended beyond the beam size. In \textbf{NGC5044}, the absorption and emission lines are most clearly visible in the CO(2-1) data. However, because this is of poor spatial resolution and we are particularly interested in the location of the gas, we made the intensity maps using the higher angular resolution CO(1-0) observation.}
    \label{fig:main_fig_2}
\end{figure*}

\begin{figure*}
	\includegraphics[width=\textwidth]{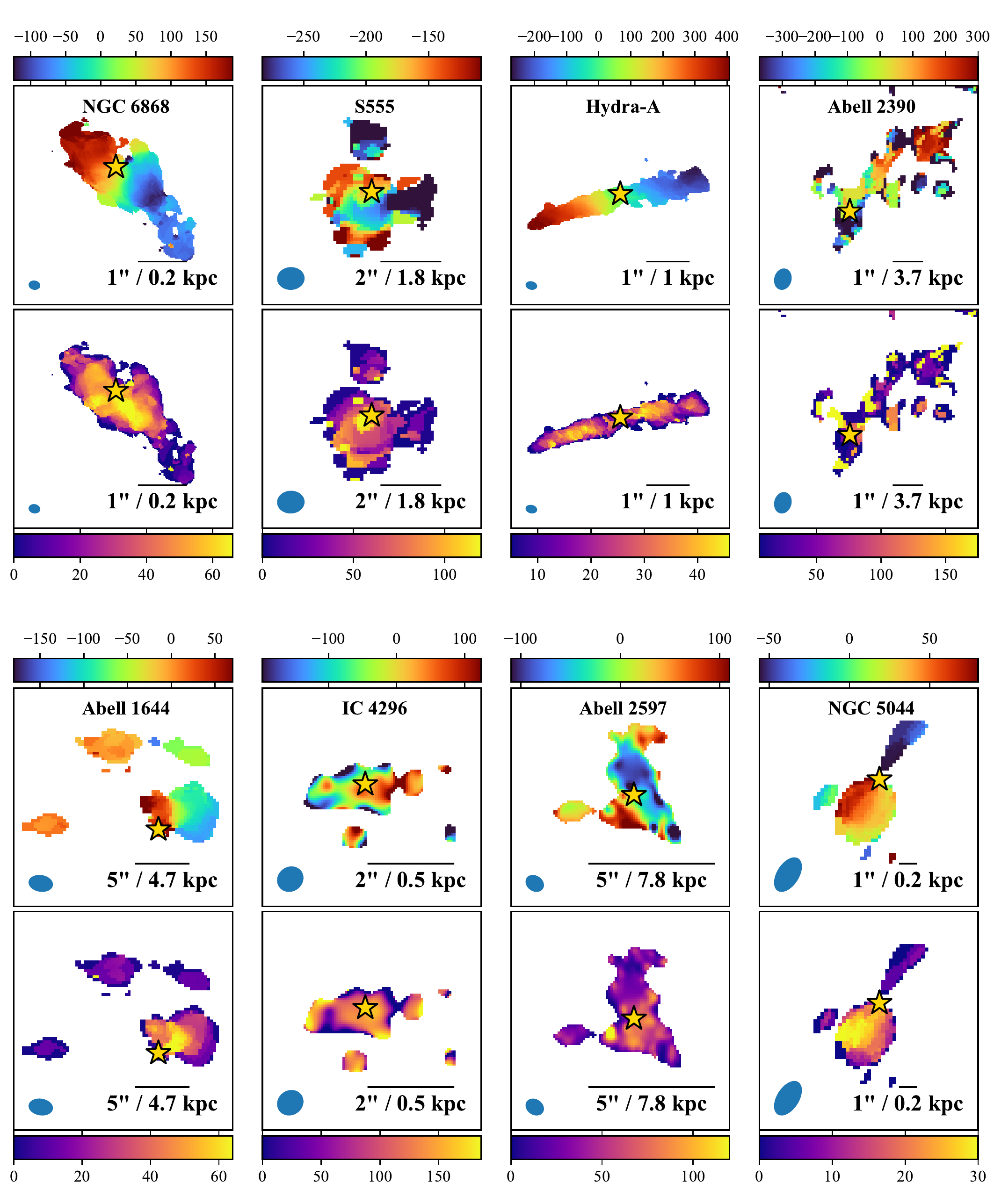}
    \caption{Velocity (upper) and dispersion (lower) maps for the eight galaxies in our sample with molecular absorption lines visible against their bright radio cores. Yellow stars indicate the position of each galaxy's radio core. All values in colour bars have units of km/s. Note that while CO(1-0) data is shown for NGC 5044 in Fig. \ref{fig:main_fig_2}, we use the stronger CO(2-1) data to make velocity and velocity dispersion maps.}
    \label{fig:velocity_and_dispersion_maps}
\end{figure*}

\section{Molecular gas properties}
\label{sec:MolecularMassesandColDensities}

In Figs. \ref{fig:main_fig_1} and \ref{fig:main_fig_2} we show the galaxies in our sample which have CO emission and absorption lines. The left half of each panel contains two spectra. One is extracted from the region over which molecular gas is visible. In some cases where the absorption is strong enough to wipe out some of the more extended emission, this extraction region excludes the radio core (these cases are made clear in the legends). The second spectrum for each source is extracted from an approximately beam-sized region centred on the radio core. This extraction region maximises the signal-to-noise ratio of the molecular absorption lines. 

The right half of each panel shows two maps of the molecular gas. One is produced using the full velocity range over which emission is visible, and the second is produced using the velocity range in which molecular absorption is present, as indicated by the grey band in the left panel. These maps are made using `includepix = [0,100]' in CASA's \texttt{immoments} task.  With these values the includepix parameter isolates flux between 0 Jy and an arbitrarily large value of 100 Jy. The cubes on which this task is performed are continuum subtracted, so any absorption takes on a negative value and is ignored by \texttt{immoments}. This prevents the absorption wiping out emission at velocities outside of the absorption region (for the sources in which the spectrum against the radio core contains both emission and absorption). Because this masks all negative values, it also has the effect of unevenly removing the noise. To account for this, the mean pixel value of the moments map is calculated in an off-center, emission free region and then subtracted from the entire image. In some cases, the moments maps on the right show that there is no detectable molecular emission along the line of sight to the galaxy's radio core with velocities matched to the absorption lines. 

In Fig. \ref{fig:velocity_and_dispersion_maps} we also show velocity and velocity dispersion maps of these galaxies made using $>3\sigma$ emission. These are made after binning the spectral cubes to $\approx 10$ km/s resolution and applying Gaussian smoothing over the beam area.

\begin{table*}
\caption{Redshifts, their corresponding velocities, velocity offsets of the molecular gas, molecular emission line integrals and molecular masses. All redshifts are barycentric and use the optical convention.}
	\centering
	\begin{tabular}{lccccr}
\hline
Source & $z_{*}$ & $v_{*}$ & $v_{*} - v_{\textnormal{mol}}$ & M$_{\textnormal{mol}}$  \\
 &  & (km s$^{-1}$) & (km s$^{-1}$) & (M$_{\odot}$)  \\
\hline
  NGC 6868 & 0.0095$\pm$0.0001 (FORS) & 2830$\pm$30 & 70$\pm40$ & 1.3$ \pm 0.2\times 10^{8}$ \\
  S555 & 0.0446$\pm$0.0001 (MUSE) & 13364$\pm$30 & -190$\pm40$ & 5.6$ \pm 0.5 \times 10^{8}$ \\
  Hydra-A & 0.0544$\pm$0.0001 (MUSE) & 16294$\pm$30 & 120$\pm50$ & 3.1$ \pm 0.2 \times 10^{9}$ \\
  Abell 2390 & 0.2304$\pm$0.0001 (VIMOS) & 69074$\pm$30 & 170$\pm40$ & 2.2$ \pm 0.6 \times 10^{10}$\\
  Abell 1644 & 0.0473$\pm$0.0001 (MUSE) & 14191$\pm$30 & -60$\pm40$ & 1.6$ \pm 0.3 \times 10^{9}$ \\
  IC 4296 & 0.01247$\pm$0.00003 (NED) & 3738$\pm$10 & -60$\pm50$ & 2.3$ \pm 0.2 \times 10^{7}$ \\
  Abell 2597 & 0.0821$\pm$0.0001 (MUSE) & 24613$\pm$30 & 20$\pm40$ & 4.6$ \pm 0.5 \times 10^{9}$ \\
  NGC 5044 & 0.0092$\pm$0.0001 (MUSE) & 2761$\pm$30 & 30$\pm40$ & 2.4$ \pm 0.4 \times 10^{7}$ \\
  RXC J1350.3+0940 & 0.13255$\pm$0.00003 (SDSS) & 39737$\pm$10 & -60$\pm30$ & 2.5$ \pm 0.3 \times 10^{10}$ \\
  MACS 1931.8-2634 & 0.35248$\pm$0.00004 (MUSE) & 105670$\pm$10 & -20$\pm30$ & 7.8$ \pm 0.7 \times 10^{10}$ \\
  RXC J1603.6+1553 & 0.10976$\pm$0.00001 (SDSS) & 32905$\pm$3 & 0$\pm30$ & 1.0$ \pm 0.1 \times 10^{10}$ \\
		\hline 
	\end{tabular}
    \label{tab:gas_properties}
\end{table*}

\subsection{Molecular masses}

The molecular mass of a galaxy with CO emission can be estimated using the following relation from \citet{Bolatto2013},

\begin{equation}
\begin{split}
\textnormal{M}_{\text{mol}} = \frac{1.05\times 10^{4}}{F_{ul}} \left( \frac{X_{\text{CO}}}{2\times 10^{20}\frac{\text{cm}^{-2}}{\text{K km s}^{-1}}}\right)\left( \frac{1}{1+z}\right) \\ \times \left( \frac{S_{\text{CO}} \Delta v}{\text{Jy km s}^{-1}}\right) \left( \frac{D_{\text{L}}}{\text{Mpc}}\right)^{2} \textnormal{M}_{\odot},
\end{split}
\label{eq:massequation}
\end{equation}
where $M_{\text{mol}}$ is the mass of molecular hydrogen, $X_{\text{CO}}$ is the CO-to-H$_{2}$ conversion factor, $z$ is the redshift of the source, $S_{\text{CO}} \Delta v$ is the CO emission integral and $D_{\text{L}}$ is the luminosity distance in Mpc. $F_{ul}$ is a factor which is included as an approximate conversion between the expected flux density ratios of the CO(1-0) and CO(2-1) lines, where $u$ and $l$ represent the upper and lower levels. For CO(1-0), $F_{10}=1$ and for CO(2-1), $F_{21}=3.2$. This value is consistent with similar studies \citep[e.g.][]{David2014,Tremblay2016}, and originates from a combination of the factor of two between the frequencies of the lines and the brightness temperature ratio observed for molecular clouds in spiral galaxies of 0.8 \citep{BraineandCombes1992}.

The CO-to-H$_{2}$ conversion factor is a considerable source of uncertainty. To ensure our mass estimates are comparable with other similar studies, we use the standard Milky Way value of \mbox{$X_{\text{CO}} = 2 \times 10^{20}$ cm$^{-2}$ (K km s$^{-1}$)$^{-1}$} in our calculations.

Table \ref{tab:gas_properties} shows the molecular gas masses of the eight galaxies shown in Figs. \ref{fig:main_fig_1} and \ref{fig:main_fig_2}. We also show the redshift and molecular gas velocity. Some of these sources have previously had their molecular gas masses estimated, but we repeat the calculations here for completeness. We also estimate the molecular masses of three brightest cluster galaxies which have emission lines but no absorption lines. These were first presented in \citet{Rose2019b}, but no analysis of them was carried out. These three galaxies are discussed later in Section \ref{sec:three_massive_bcgs}. 

 The stellar redshifts of Hydra-A, Abell 1644 and NGC 5044 are taken from Multi Unit Spectroscopic Explorer (MUSE) observations (ID: 094.A-0859). The MUSE stellar redshift we use for Abell 2597 is consistent with the value from Ca \textsc{II, H+K} absorption lines \citep{Voit1997}. The stellar redshifts of RXC J1350.3+0940 and RXC J1603.6+1553 are derived from a combination of emission and absorption lines detected with the the Sloan Digital Sky Survey (SDSS) \citep{SLOAN}. The stellar redshift of MACS 1931.8-2634 is taken from \citet{Fogarty19} and is found using MUSE observations. Crosschecking with FOcal Reducer and low dispersion Spectrograph (FORS) observations of S555, Abell 1644, NGC 5044, and Abell 2597 provides redshifts in good agreement with those listed. The redshifts used for Abell 2390 and RXCJ0439.0+0520 are taken from Visible Multi-Object Spectrograph (VIMOS) observations previously presented by \citet{Hamer2016} and are based primarily on stellar emission lines. The observed wavelengths of the single stellar absorption line in these two VIMOS spectra are consistent with the quoted redshifts. The redshifts we use here are the same as those given in \citet{Rose2019b}. The exception to this is IC 4296, which was not presented in that paper. For this source we have used the same value as \citet{Ruffa2019}.)

To estimate the molecular masses, we first make Gaussian fits to the emission lines extracted from the region over which molecular gas is visible. Depending on the profile of the emission, we use either a single Gaussian or double Gaussian line. For sources with absorption lines which are embedded within (or are near to) the emission, we mask the absorption before making the fits. The masked velocity range is the same as that shown by the grey band in Figs. \ref{fig:main_fig_1} and \ref{fig:main_fig_2} (except in IC 4296, where we mask the range between the grey bands).

For most sources estimating the molecular mass is a straightforward process. However, A2390 and NGC 5044 require more detailed explanation. 

A significant amount of emission in A2390 is in close proximity to the radio core. The observation also has a relatively large beam size and a wide molecular absorption line. If a spectrum is extracted from the entire region over which molecular emission is present, the absorption line wipes out the spatially extended molecular emission which is shown by the red line in Fig. \ref{fig:main_fig_1}. This results in a severely underestimated molecular mass. The molecular emission from the core region has a wider velocity range, and is still visible at velocities below that of the absorption (see the blue line in Fig. \ref{fig:main_fig_1}). We therefore extract spectra from two non-overlapping regions: one surrounding the radio core, and another which is more spatially extended. The masses from these two regions are then calculated following the procedure outlined above and summed together.

Molecular emission in NGC 5044 has been detected in both CO(1-0) and CO(2-1). The CO(2-1) line is the stronger of the two and has been analysed by \citet{David2014, Temi2018, Schellenberger2020}. Nevertheless, in Table \ref{tab:gas_properties} we present a new mass estimated from the spatially resolved CO(1-0) emission. We find a mass of $2.4\pm 0.4 \times 10^{7} \textnormal{M}_{\odot}$ from a 6x2" rectangular extraction region centred on the continuum and at an angle of 40 degrees. This is similar to the value of $4.2\pm0.1 \times 10^{7} \textnormal{M}_{\odot}$ found by \citet{Schellenberger2020}, from a circular radius of 6". No CO(1-0) emission is clear from outside the rectangular extraction region used (see Fig. \ref{fig:main_fig_2}), but for completeness we also estimate the mass from within the 6" radius used by \citet{Schellenberger2020} on the CO(2-1) data. This gives a mass of $3.5\pm 0.6 \times 10^{7} \textnormal{M}_{\odot}$, so some weak emission may in fact be present outside the rectangular region described. NGC 5044 also contains more spatially extended molecular gas visible to the ALMA Compact Array (ACA). From ACA observations, \citet{Schellenberger2020} find a mass of $6.5 \pm 0.2 \times 10^{7} \textnormal{M}_{\odot}$ within a 15" radius.

\subsection{Three brightest cluster galaxies with $> 10^{10} \textnormal{M}_{\odot}$ of molecular gas}
\label{sec:three_massive_bcgs}

\begin{figure*}
	\includegraphics[width=\textwidth]{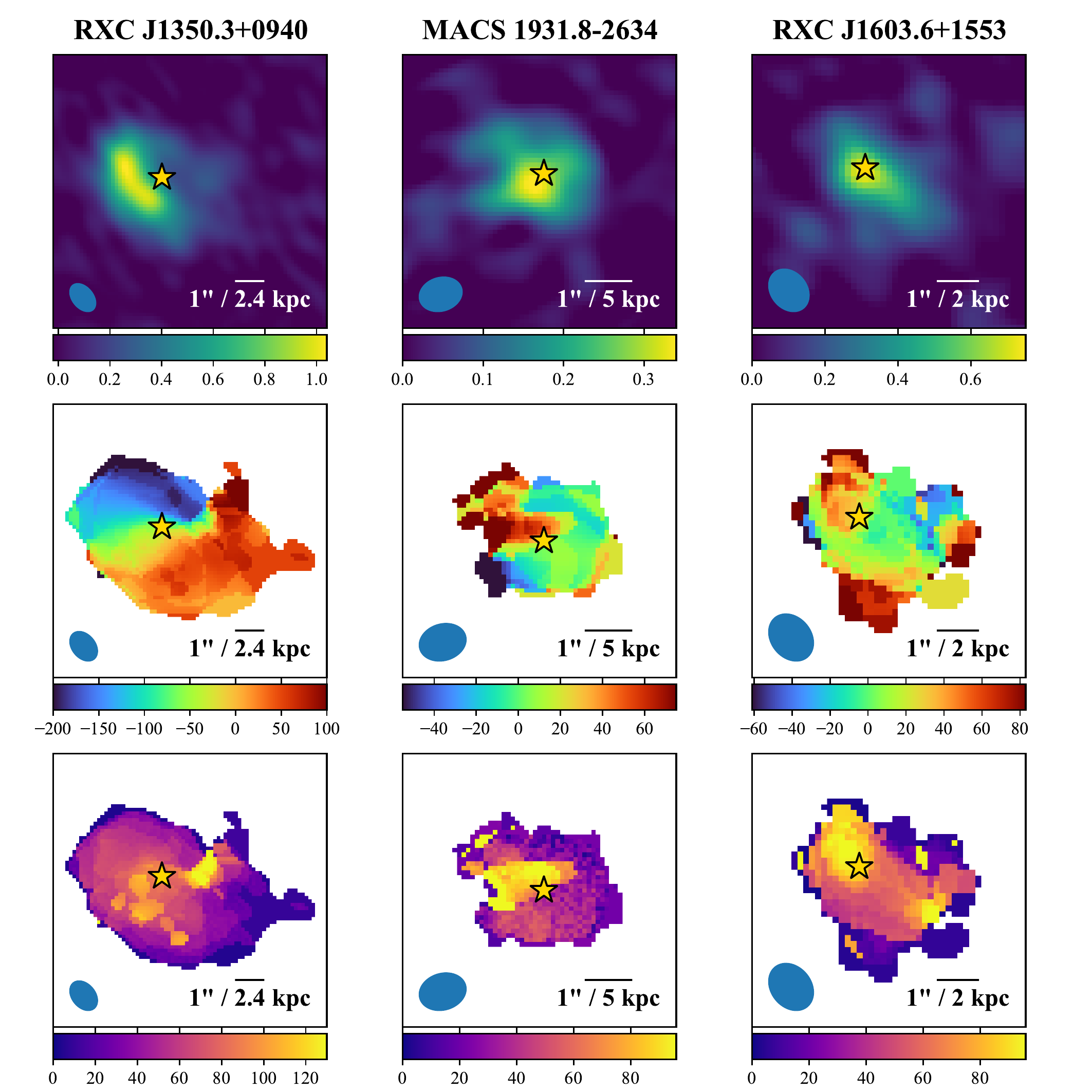}
    \caption{From top to bottom: Intensity, velocity and velocity dispersion maps for three brightest cluster galaxies with $>10^{10}\textnormal{M}_{\odot}$ of molecular gas, but no absorption. Units are Jy\,km/s, km/s and km/s.}
    \label{fig:maps_of_emitters}
\end{figure*}

The largest single survey searching for molecular absorption lines in brightest cluster galaxies was completed in ALMA Cycle 6 (project 2017.1.00629.S, PI Edge). In total, 20 sources were observed and eight had detections of molecular absorption against their radio cores \citep[][]{Rose2019b}. Several of the targets from this survey in which molecular absorption was found are shown in Figs. \ref{fig:main_fig_1} and \ref{fig:main_fig_2}.

In three of the targets molecular emission was detected, but no absorption lines were found. We introduce them here and briefly discuss the properties of their molecular emission. Since these targets were observed as part of the same survey in which most of the absorbing sources were identified, they make for an interesting comparison.

In Fig. \ref{fig:maps_of_emitters} we show their intensity, velocity and velocity dispersion maps. Below we briefly summarize the properties of their molecular emission. 

\begin{itemize}
\item \textbf{RXC J1350.3+0940} has $2.5 \pm 0.3 \times 10^{10} \textnormal{M}_{\odot}$ of molecular gas in a slightly inclined edge-on disc with a smooth velocity gradient, a velocity range of approximately 300 km/s and diameter of 10 kpc.
\item \textbf{MACS 1931.8-2634} has a molecular mass of $7.8 \pm 0.7 \times 10^{10} \textnormal{M}_{\odot}$ centred a few hundred parsecs from its radio core. There is no clear velocity structure. The gas has a roughly circular distribution with a diameter of approximately 5 kpc. The molecular gas of MACS 1931.8-2635 has previously been studied by \citet{Fogarty19}, though the CO(1-0) observation we present has not previously been published. With the Galactic $X_{\textnormal{CO}}$ conversion factor which we use throughout this paper, \citet{Fogarty19} find a molecular mass of $9.4 \pm 1.3 \times 10^{10} \textnormal{M}_{\odot}$ from their CO(1-0) observation.
\item \textbf{RXC J1603.6+1553} has $1.0 \pm 0.1 \times 10^{10} \textnormal{M}_{\odot}$ of molecular gas, with the brightest emission centred on its radio core. The gas is weakly resolved spatially but appears to extend across a diameter of around 5 kpc. There is no clear velocity structure. 
\end{itemize}

The molecular masses in these systems are extremely high and make them among the most gas rich systems known \citep[][]{Pulido2018}. All three also have strong emission coincident with their radio cores, so their lack of absorption features is surprising. This is particularly so given that galaxies with molecular masses lower by approximately two orders of magnitude (NGC 6868 and S555) have clear CO(1-0) absorption. NGC 5044 also has absorption, despite having no detectable emission along the line-of-sight to its radio core.

Therefore, the probability of seeing molecular absorption against a galaxy's radio core is not strictly dependent on the total amount of molecular gas present, but more so on the molecular column density and gas properties (such as the excitation temperature).

\subsection{Column density estimates from emission}

The blue spectra in Figs. \ref{fig:main_fig_1} and \ref{fig:main_fig_2} are extracted from compact regions centred on each galaxy's continuum source. In all these spectra, emission is present. This suggests there may be a significant amount of molecular gas along the line-of-sight to the continuum.

However, the continuum sources are significantly smaller than the angular resolution of the observations. Therefore, although the emission and absorption in these spectra are extracted from the same region, they originate on different spatial scales. Typically, the beam sizes are of arcsec scale, while continuum sources are of milliarcsec scale. As a result, any emission in these spectra will come from a physically much larger region than the absorption, so the two are not directly comparable. 

The emission strength is also heavily dependent on factors such as redshift and beam size, so it is an unreliable indicator of how much gas is likely to be covering the continuum. However, these parameters are known accurately, so the average column density of emitting gas covering the continuum source can be found by assuming a uniform density within the beam. This then serves as an estimate of the column density of molecular gas covering the continuum source \citep[unless there is a significant amount of molecular gas which is too cool to be visible in emission, as suggested by][]{Fabian2022}. The interstellar medium is clumpy, so this calculation will give an average column density across many different lines of sight. In reality, some lines of sight may intersect many clouds and have a higher column density. Other lines of sight may intersect negligible amounts of gas.

When using a spectrum extracted from a beam sized region centred on the continuum source, the column density of molecular gas is

\begin{equation}
    \sigma_{\text{mol}} = \frac{\textnormal{M}_{\text{mol}}}{2\textnormal{A}_{\textnormal{beam}}}\,\textnormal{M}_{\odot}/\textnormal{pc}^{2},
\end{equation}

where $\textnormal{M}_{\text{mol}}$ is the mass in M$_{\odot}$ estimated from equation \ref{eq:massequation}, and $\textnormal{A}_{\textnormal{beam}}$ is the beam area in pc$^{2}$. The factor of 2 accounts for only gas on one side of the galaxy being able to cover the continuum source, whereas the emission comes from gas on both sides.

In Table \ref{tab:gas_densities} we show the column density of molecular gas expected to cover each continuum source. Most of the estimated values are similar to the typical mass and column densities of giant molecular clouds. \citet{Lombardi2010} find typical masses of 170 to 710\,$\textnormal{M}_{\odot}$ for giant molecular clouds, while their column densities are found to be 50\,$\textnormal{M}_{\odot}/\textnormal{pc}^{2}$ in the Large Magellanic Cloud \citep[][]{Hughes2010} to 170\,$\textnormal{M}_{\odot}/\textnormal{pc}^{2}$ in the Milky Way \citep[][]{Solomon1987}. In turn, this implies a low covering fraction of molecular clouds \citep[as is also predicted by simulations, e.g.][]{Gaspari2017}. Therefore, as previously stated, clumpiness may mean some lines-of-sight to the continuum intersect several molecular clouds, while others intersect none.

\begin{table}
\caption{The column density of molecular gas covering each galaxy's continuum source, estimated from emission and absorption seen along the line-of-sight. Due to uncertainties in constants used in calculations, all values carry an error of approximately 50 per cent. \newline $^{1}$The emission coincident with the continuum source in Abell 1644 is unresolved by an observation with a relatively large beam size, so we mark this column density as a lower limit.}
	\centering
	\begin{tabular}{lccc}
	
\hline
& \multicolumn{2}{c}{Column Density} & Clumping Factor \\
& Emission & Absorption \\
\hline
NGC 6868 & 960 & 1100 & 1.1 \\
S555 & 130 & 1300 & 9.8\\
Hydra-A & 530 & 210 & 0.4 \\
Abell 2390 & 160 & 15000 & 92 \\
A1644$^{1}$ & >50 & 310 & <6\\
IC 4296 & 170 & 290 & 1.1 \\
Abell 2597 & 130 & 6800 & 54 \\
NGC5044 & <50 & 3800 & >76 \\
RXC J1350.3+0940 & 130 & <700 & <5.4 \\
MACS 1931.8-2634 & 240 & <2200 & <9.5 \\
RXC J1603.6+1553 & 330 & <3400 & <10 \\
\hline
	\end{tabular}
    \label{tab:gas_densities}
\end{table}

\subsection{Column density estimates from absorption}
%https://ned.ipac.caltech.edu/level5/Sept13/Bolatto/Bolatto4.html

The line of sight column density, $N_{\textnormal{tot}}$, of an optically thin molecular absorption region is:
\begin{equation}
\label{}
N_{\textnormal{tot}}^{\textnormal{thin}} = Q(T_{\textnormal{ex}}) \frac{8 \pi \nu_{ul}^{3}}{c^{3}}\frac{g_{l}}{g_{u}}\frac{1}{A_{ul}} \frac{1}{ 1 - e^{-h\nu_{ul}/k T_{\textnormal{ex}}}}\int \tau_{ul}~dv~,
\label{eq:thin_colum_density}
\end{equation}
where $Q$($T_{\textnormal{ex}}$) is the partition function, $c$ is the speed of light, $A_{ul}$ is the Einstein coefficient of the observed transition and $g$ the level degeneracy, with the subscripts $u$ and $l$ representing the upper and lower levels \citep{Godard2010,Magnum2015}.

The assumption of optically thin absorption in equation \ref{eq:thin_colum_density} is inappropriate for some of our data. To account for this, a correction factor can be applied \citep{Magnum2015} to give the following more accurate column density,
\begin{equation}
N_{\textnormal{tot}} = N_{\textnormal{tot}}^{\textnormal{thin}} \frac{\tau}{1-\exp({-\tau})}.
\label{eq:true_column_densities}
\end{equation}{}

Equation \ref{eq:thin_colum_density} gives the column density of the absorbing molecule as a number density per square cm. To make these column densities comparable with those estimated from the emission in the previous subsection, we convert this to a column density in units of $\textnormal{M}_{\odot}/\textnormal{pc}^{2}$. This requires the assumption of an H$_{2}$/CO number ratio, which we assume to be constant across all of our sources. This ratio has been estimated using optically thin isotoplogues, such as $^{13}$CO. Observations of the Milky Way's Taurus molecular cloud by \citet{Pineda2010} give an estimate of CO/H$_2$ = 1.1$\times 10^{-4}$ (with an assumed isotopic ratio of CO/$^{13}$CO = 70). Similarly, \textit{Far Ultraviolet Spectroscopic Explorer} observations by \citet{Rachford2009} give a ratio of CO/H$_2$ = 1.8$\times 10^{-4}$. In our calculations, we use a ratio of CO/H$_2$ = 1.5$\times 10^{-4}$. We also assume an excitation temperature of 5K. This is the approximate value for the absorbing clouds in Hydra-A \citep[][]{Rose2020}.

Table \ref{tab:gas_densities} shows the column densities estimated using each object's molecular absorption lines. To calculate these we use the most optically thin line available in each case e.g. the weaker CO(1-0) line in Hydra-A from \citet[][]{Rose2019a}. We also show upper limits for the absorption column densities of the three sources from the \citet[][]{Rose2019b} survey which have molecular emission but no detectable absorption. These upper limits are found assuming an absorption line with a velocity dispersion of $\sigma = 20$ km/s.

\section{Discussion}

\subsection{Two types of molecular absorption line system}

In section \ref{sec:MolecularMassesandColDensities}, we employed two methods to estimate the column density of molecular gas along the line-of-sight to each object's continuum source. First we used the molecular emission, which indicates the amount of gas from the galaxy-scale distribution which is expected to cover the continuum. Secondly, we derived the column density from the absorption. This gives the actual amount of molecular gas which covers the continuum. 

The clumping factors we find range from 0.4 (where just less than half as much absorbing gas is seen as would be expected on an average line of sight) to 92 (where 92 times as much gas is seen in absorption as would be expected on an average line of sight). We find no correlation between the clumping factor and the beam size or redshift. It is also worth noting that the clumping factors we find are heavily dependent on the excitation temperature and CO-to-H$_{2}$ conversion factor used. However, unless there is a large and unexpected variation in these values between individual sources, comparison of their relative values is meaningful.

That having been said, if each source's absorption is caused by the same population of clouds as are responsible for the emission, the larger clumping factors imply that the absorption lines are tracing significantly more molecular gas than expected in these systems. Alternatively, the high variance may be explained by the clumpiness of the interstellar medium's molecular gas.

\begin{figure*}
	\includegraphics[width=\textwidth]{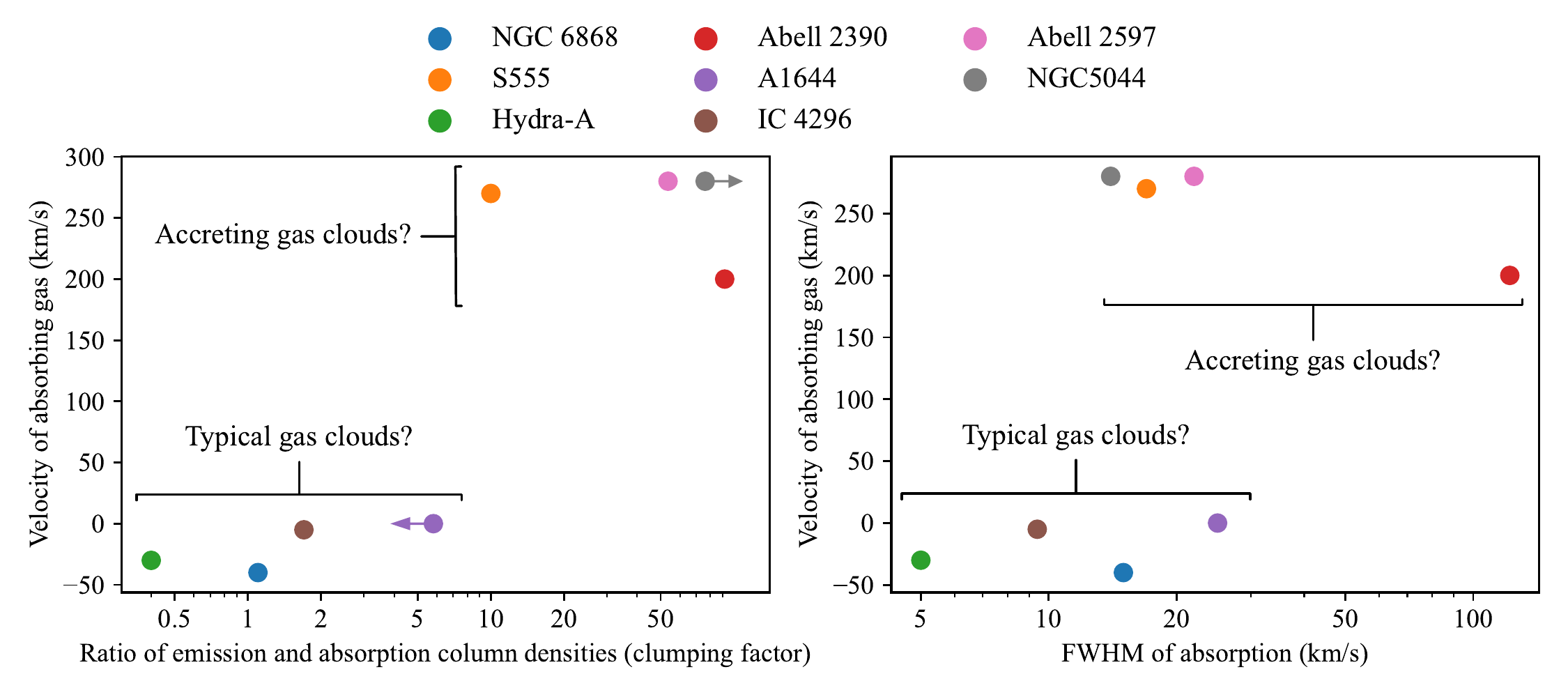}
    \caption{The clumping factor and FWHM of the absorbing gas versus its central velocity. Positive velocities mean inward motion towards the centre of the galaxy. The clumping factor is the ratio of two column density estimates derived from (i) the molecular emission from a beam-sized region spatially coincident with the radio core, and (ii) the molecular absorption against the radio core. A clumping factor of 1 implies that the absorption is tracing a column density equal to the average amount of emitting gas covering a line of sight to the continuum source. A clumping factor of 2 implies that twice as much gas is being traced. In the right panel, for sources with multiple absorption features we use the FWHM of the deepest.}
    \label{fig:detection_factors}
\end{figure*}

In Fig. \ref{fig:detection_factors} we show the clumping factor of each source (and the absorption's FWHM) plotted against the mean absorption velocity. Two distinct populations are apparent. The first have clumping factors close to unity (narrow FWHM) and the absorbing gas is moving at close to 0 km/s. The velocity of the absorbing gas is also similar to that of the molecular emission. The second population have high clumping factors (wide FWHM) and the absorbing gas has high velocities towards the galaxy core. These velocities are dissimilar to that of the molecular emission (except in Abell 2390, where the emission is unusually wide at 900 km/s).

In the following sections, we argue that these two populations represent two distinct types of absorption system. The first contain absorption due to chance alignments between the continuum source and the galaxy-wide distribution of molecular gas -- the same gas which is responsible for the galaxy's molecular emission. The second population contain absorption which is due to a distinct and much less numerous set of molecular clouds which are in the process of accreting onto the AGN. Due to the scarcity these clouds, they are not visible in emission, but their proximity to the galaxy centre makes alignment with the continuum more likely.

\subsection{The expected properties of molecular absorption lines}

We have proposed that the absorption lines in Figs. \ref{fig:main_fig_1} and \ref{fig:main_fig_2} are due to two distinct populations of molecular gas clouds -- those spread throughout the galaxy and those at lower radii which are accreting onto the AGN. We now explain how the properties of the absorption caused by these two clouds populations arise.

\subsubsection{Absorption due to clouds visible in emission (typical clouds)}

A CO spectrum seen against the central continuum source of a galaxy can contain the following components (i) emission, produced by $10^{7}$M$_{\odot}$ or more of molecular gas lying across a region of several hundred parsecs, and (ii) absorption, produced by approximately several hundred to several thousand solar masses of molecular gas which lies along the narrow line-of-sight to the galaxy's bright radio core \citep[e.g.][]{Rose2020}.

A significant column density is required for a fraction of the gas spread across a galaxy's disc to have a high probability of occulting the continuum and producing molecular absorption. Typically, molecular clouds in the Milky Way are clumpy and have column densities of around 170 M$_{\odot}$/pc$^{2}$ \citep[][]{Solomon1987}. Therefore, for a continuum source a few square parsec in size, a column density of a few hundred M$_{\odot}$/pc$^{2}$ would be required for an alignment between the continuum and one or more molecular clouds to be probable -- depending on the level of gas clumping. If there is an alignment, the velocity dispersion of a single cloud's absorption would likely be similar to that of molecular clouds in the disc of the Milky Way, i.e. rarely above 8 km/s \citep[][]{Miville2017}. It would also be similar to the velocity dispersion found for molecular clouds detected in galaxy discs when using a high redshift quasar as an absorption backlight, i.e. also rarely above 8 km/s \citep[e.g.][]{Wiklind1996a,Combes2008, Muller2013}. However, multiple closely associated molecular clouds with similar velocities may combine to produce a wider absorption feature.

Absorption due to chance alignments between the continuum source and a fraction of the clouds which are visible in emission will exist along an extremely narrow line of sight. However, the beam of any observation will be much larger than this and may contain enough molecular gas to be visible in emission. In these cases, the emission spectrum (i.e. velocity vs. flux) can be interpreted as showing the relative number of clouds as a function of velocity. Clouds with a velocity at the peak of the emission are the most numerous, so clouds at these velocities would be the most likely to intersect the continuum. Absorption at or close to the peak of the emission is therefore most likely in this situation. This will also be true for any absorption lines produced by extremely cool gas not visible in emission, assuming the cooler and warmer gas are dynamically linked.

Throughout the rest of this paper, we use the term `typical' in reference to those clouds which make up a galaxy's molecular emission.

\subsubsection{Absorption due to clouds close to the AGN}

Absorption due to molecular clouds close to the centre of a galaxy are not forbidden from having similar characteristics to those expected of more typical clouds which are at larger radii, but in all likelihood their line-of-sight velocity and velocity dispersion will be different. 

The relative probability of an absorption line being produced by a cloud in the core of a galaxy is considerably higher than it is for one further out. As simulations \citep[such as][]{Gaspari2017, Krumholz2016} show, the volume filling factor and internal density of molecular clouds is inversely proportional to their distance from the galaxy centre. Since these clouds are more likely to occult the continuum than those at larger radii, they won't necessarily be sufficiently numerous to make them visible in emission. Therefore, the absorbing clouds need not have a similar velocity to the clouds visible in emission. Instead, due to their presence in the strong gravitational field, they can attain high inward velocities and be redshifted compared with the galaxy's emission.

A second way in which absorption from gas near the galaxy centre may differ is in its width. The nuclear region of a galaxy is a hostile environment with a large velocity gradient. This would quickly be imprinted onto any gas cloud(s), so the absorption would likely have a higher velocity dispersion than if it was produced by gas clouds at larger radii in the galaxy, i.e. $\gtrsim$8\,km/s. By the same argument, they would have a higher velocity dispersion than absorption lines seen at radii of several hundred pc or more in galaxies where a high redshift quasar has been used as the backlight for absorption \citep[e.g.][]{Wiklind1996a,Combes2008, Muller2013}. 

\subsection{Where is the absorbing gas located in each galaxy?}
\label{sec:whereisabsorbinggas}

We have suggested the existence of two types of absorption system. When considering the molecular emission and absorption properties, these two types of absorber are indicated by the line-of-sight velocities, column densities and velocity dispersions. 

In some cases, this seems to concur with claims that these observations are tracing AGN accretion. Where the absorbing gas has inward velocities of several hundreds of km/s, ongoing supermassive black hole accretion has been suggested, e.g. A2597 by \citet{Tremblay2016}, S555 and A2390 by \citet{Rose2019b}. For a molecular gas cloud at a radius of 100\,pc and with an average inward velocity of 500 km/s, the supermassive back hole could be reached in approximately $2\times 10 ^{5}$ years. This is short on galactic timescales, so if the clouds really are at these distances from the centre of their host galaxy, their accretion is a plausible end result.

We now consider the properties of each absorption system in detail to more tightly constrain the location of the absorbing gas.

\begin{itemize}
    \item \textbf{NGC 6868}'s molecular gas is centred on the radio core, extends over approximately 0.5 kpc, and has a mass of $1.3 \pm 0.2 \times 10^{8}\textnormal{M}_{\odot}$. The gas appears to be in a disc viewed at a small angle of inclination (perhaps $10-20^{\circ}$) and has a velocity range of approximately 300 km/s. We are likely seeing typical clouds in this galaxy (i.e. those away from the core and in the more extended disc) for three reasons. Firstly, the column density of molecular gas along the line-of-sight to the continuum is relatively high at 960\,M$_{\odot}$/pc$^{2}$, so an alignment between the continuum and several of its molecular clouds is expected. This expected column density is also very similar to the actual column density of the absorption (1100\,M$_{\odot}$/pc$^{2}$). When chance alignments between the continuum and molecular clouds visible in emission are responsible for the absorption, they would likely have velocities which match the emission, as is the case. The absorption regions also have relatively low velocity dispersions ($\sigma = 6.4$ km/s, FWHM = 15 km/s), typical of clouds within the disc of a galaxy.
    \item \textbf{S555}'s molecular gas is poorly resolved and no velocity structure is discernible. However, the gas is heavily concentrated and centered around 200\,pc from the radio core, with weak emission extending out to approximately 3\,kpc. The total mass is $5.6\pm0.2\times 10^{8}\textnormal{M}_{\odot}$. In this case, the properties of the absorption indicate that it is not due to the same population of clouds responsible for the emission. The molecular column density is 130\,M$_{\odot}$/pc$^{2}$ for the gas causing the emission, so the probability of an alignment is small but not insignificant. However, the actual column density of the absorbing gas is 10 times higher than this, and no molecular emission is found at velocities matched to the absorption lines. Indeed, any absorption which does come from the typical gas clouds would be embedded within the emission -- at least 200 km/s below where the absorption is actually present. The gas is therefore moving towards the galaxy centre at high velocity, and a velocity which is significantly different to the gas seen in emission. On the other hand, the velocity dispersion of the absorption is only moderate ($\sigma = 7.2$ km/s, FWHM = 17 km/s). This is lower than might be expected if the clouds were subject to the high velocity gradient close to a supermassive black hole. On the balance of probabilities, the absorbing gas is most likely near the galaxy centre and in the process of accretion. However, due to its moderate velocity dispersion it is unlikely to be in the very centre of the galaxy.
    \item \textbf{Hydra-A} has a 5 kpc wide edge-on molecular gas disc with a total mass of $3.1\pm0.2 \times 10^{9}\textnormal{M}_{\odot}$, a smooth velocity gradient and a velocity range of approximately 700 km/s \citep[][]{Rose2019a}. Lower angular resolution CO(2-1) observations from IRAM place the mass at $2.26 \pm 0.29 \times 10^{9}\textnormal{M}_{\odot}$ \citep[][]{Hamer2014}. The molecular gas with velocities matched to the absorption is centred on the radio core (Fig. \ref{fig:main_fig_1}) and extended by a few hundred parsecs along the blueshifted edge of the disc. The column density of molecular gas against the radio core is 530\,M$_{\odot}$/pc$^{2}$, similar to the 210\,M$_{\odot}$/pc$^{2}$ derived from the absorption. Estimates of the diameter of the continuum source in Hydra-A have been placed at between 4\,pc and 7\,pc, so on average we would expect approximately a few tens of $\sim170$ M$_{\odot}$ clouds from the galaxy's disc to lie along the line of sight. Hydra-A is by far the best studied absorption system of this kind, and this matches the observations of \citet{Rose2020}. They identified at least 12 molecular clouds which are responsible for Hydra-A's absorption. The absorbing clouds are also at the peak of the emission profile -- the most likely velocity at which we would detect clouds lying in the disc. Additionally, the velocity dispersions of the individual clouds are very low (<5 km/s, FWHM <12 km/s). In terms of the number of clouds, their velocities, and velocity dispersions, the absorption in Hydra-A therefore has the properties expected as a result of chance alignments between the continuum and gas clouds in the galaxy's disc.
    \item \textbf{Abell 2390} contains a significant molecular mass of $2.2\pm0.6 \times 10^{10}\textnormal{M}_{\odot}$, though due to its relatively high redshift the emission is weak and poorly resolved. Nevertheless, the molecular gas is significantly extended across a strongly asymmetric filamentary structure approximately 20 kpc long and 5 kpc wide. Evidence of velocity structure to the emission is observed, with higher velocities being found at larger radii (Fig. \ref{fig:velocity_and_dispersion_maps}), perhaps indicating a molecular outflow. Abell 2390's molecular emission implies an average column density of molecular gas along the line-of-sight to the radio core of 160\,M$_{\odot}$/pc$^{2}$, so the probability of several of the clouds we see in emission aligning with the continuum is small -- unless the line of sight passes through a particularly dense clump of the ISM. At 15000\,M$_{\odot}$/pc$^{2}$, the absorption in Abell 2390 is 92 times stronger than the emission would predict. The absorption also has a dispersion of 52 km/s \citep[FWHM = 120 km/s,][]{Rose2019b} -- extremely wide considering it is produced by just 15000\,M$_{\odot}$/pc$^{2}$ of gas. That implies the absorbing gas is in a region with a very high velocity gradient. Such a gradient can occur within a few parsecs of a supermassive black hole. However, the emission peak from gas on large scales closely matches the absorption line speeds, which may be more than coincidence. The broad, 800 km/s emission seen towards the nucleus indicates a second emission component that may be associated with the nucleus or an outflow. This complexity prevents a conclusive statement, but due to the high velocity dispersion and inflow speed, the absorbing gas is probably close to the core and accreting onto the AGN.
    \item \textbf{Abell 1644} has $1.6\pm0.3 \times 10^{9}\textnormal{M}_{\odot}$ of molecular gas, the bulk of which is contained in four main clumps distributed along an arc extending out from the galaxy centre. This arc of gas has been spatially matched to X-ray emission, suggesting it has cooled from the hot intracluster medium \citep[][]{Baek2022}. Two of the gas clumps are within approximately 200\,pc of the radio core, while the others are at radii of approximately 7 and 10 kpc. The absorption in Abell 1644 is close to the peak of the spatially matched emission. This implies that the absorption is due to a sample of the same clouds responsible for the emission. The velocity dispersion is relatively high, ($\sigma = 11$ km/s, FWHM = 26 km/s), though from more recent ALMA observations of HCO$^{+}$, we know this is due to it containing multiple components (Rose et. al, in prep.). The column density of molecular gas along the line-of-sight to the radio core has a lower limit of >50 M$_{\odot}$/pc$^{2}$, so we are unable to say how much gas from the emitting population would be expected to lie along the line-of-sight (this is a lower limit because the molecular emission coincident with the radio core is unresolved in a relatively large beam). Nevertheless, even the upper limit this gives for the clumping factor (<6) implies that the absorption is not tracing significantly more gas than the emission would predict. This, combined with the line-of-sight velocity and velocity dispersion of the absorption implies that the absorption in Abell 1644 is caused by typical clouds within the galaxy.
    \item \textbf{IC 4296} has a molecular mass of $2.3\pm 0.2\times 10^{7}\textnormal{M}_{\odot}$ in a disc-like structure approximately 0.8 kpc wide and with a weak velocity gradient \citep[][find a molecular mass of $2.0\pm 0.2\times 10^{7}\textnormal{M}_{\odot}$]{Ruffa2019}. The angular resolution of the observation is poor, so only an upper limit of 0.2 kpc can be placed on the disc's width. Emission with velocities matched to the absorption is spread across most of the disc-like structure, except at its extremities. The properties of the molecular absorption are entirely consistent with it being due to clouds within the disc of the galaxy. The covering factor of the continuum is 170\,M$_{\odot}$/pc$^{2}$, similar to the actual column density of the absorption along the line of sight of 290\,M$_{\odot}$/pc$^{2}$. The absorption also has a narrow velocity dispersion (for the deepest line, $\sigma = 4$ km/s, FWHM = 9.4 km/s) and the absorbing clouds lie at the peak of the molecular emission.
    \item \textbf{Abell 2597} has an approximately 10 kpc wide concentration of molecular gas centred on its radio core, with a velocity range of 200 km/s. More diffuse emission extends 30 kpc out to the north and south, taking the total molecular mass to $4.6\pm0.5 \times 10^{9}\textnormal{M}_{\odot}$ \citep[][find a mass of $3.2 \pm 0.1\times 10^{9}\textnormal{M}_{\odot}$. From the extraction region they show, this mass likely includes less of the extended emission]{Tremblay2018}. In Figs. \ref{fig:main_fig_2} and \ref{fig:velocity_and_dispersion_maps}, we only show the nuclear region, but wider maps can be seen in \citet{Tremblay2018}. 
    Molecular emission with velocities matched to the absorption lies outside the radio core, at radii between 300\,pc and 3\,kpc. The column density of molecular gas towards the continuum is estimated at 130\,M$_{\odot}$/pc$^{2}$, so on average only a small number of clouds would be expected to cover the continuum. This is in stark contrast to the absorption's actual column density of 6800\,M$_{\odot}$/pc$^{2}$, suggesting it is not caused by the same population of clouds responsible for the emission. The line-of-sight velocities of the absorbing gas clouds are also dissimilar to those causing the molecular emission. Furthermore, the width of the three resolved absorption features is moderately high (roughly $\sigma = 9$ km/s, FWHM = 21 km/s for each component). This is higher than is typically seen for absorption due to molecular clouds at several hundreds of parsec or more from their galaxy centre \citep[e.g.][]{Wiklind1996a, Muller2013}. All the properties of the absorption in this system suggest the gas is most likely in the central regions of the galaxy. Due to the high velocity towards the core, accretion can be inferred.
    \item In \textbf{NGC5044}, the highest spatial resolution observations of the molecular gas are with CO(1-0), and suggest a mass of $2.4\pm0.4 \times 10^{7}\textnormal{M}_{\odot}$ \citep[with CO(2-1)][find a mass of $4.2\pm0.1\times 10^{7}\textnormal{M}_{\odot}$]{Schellenberger2020}. The weak CO(1-0) emission is offset from the continuum by around 300pc. An accurate estimate of its diameter is hard to attain due to the weakness of the emission, but it may be up to 1.5\,kpc. No emission is detectable along the line-of-sight to the radio core, so only a <50\,M$_{\odot}$/pc$^{2}$ upper limit for the column density of emitting gas can be found, much lower than the actual 3800\,M$_{\odot}$/pc$^{2}$ column density of the absorbing gas. This, combined with the absorption being at least 100 km/s offset from the emission implies that the absorbing clouds are not drawn from the population visible in emission. Instead, they are more likely close to the galaxy centre where their probability of intersecting the continuum is relatively high. However, although the absorbing gas is moving inwards at 280 km/s and is likely close to the galaxy centre, we cannot definitively confine it to the immediate surroundings of the radio core because its velocity dispersion ($\sigma = 5.9$ km/s, FWHM = 14 km/s) is lower than might be expected in this region.
\end{itemize}

\subsection{Constraining the location of typical absorbing clouds}

\begin{figure*}
	\includegraphics[width=\textwidth]{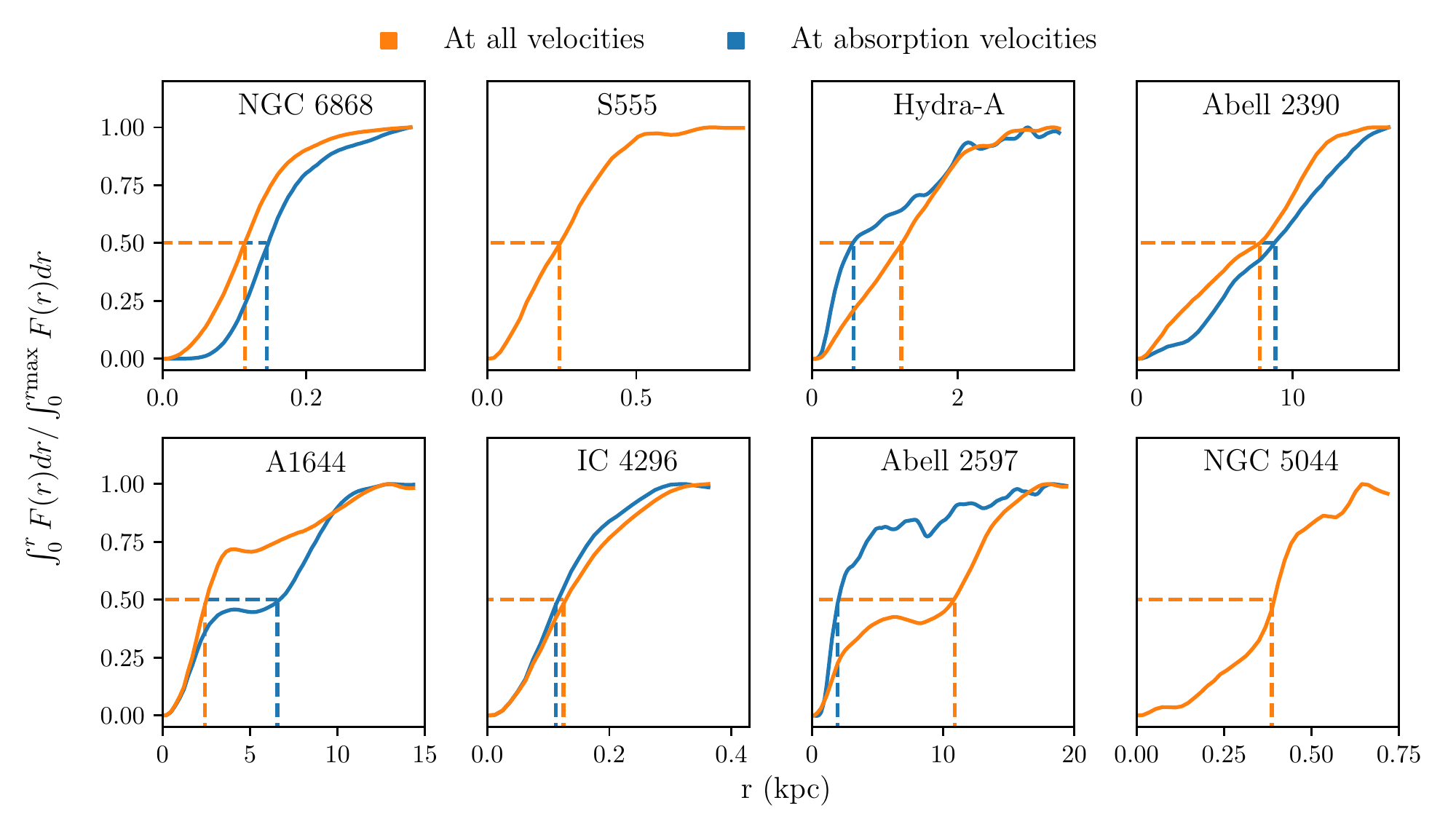}
    \caption{The fraction of the CO flux contained within a given radius, for flux at all velocities (orange) and velocities which match that of the absorption regions (blue). Blue and orange dashed lines indicate the half-light radii. This effectively gives an upper limit for the probability that a randomly selected gas cloud will lie within a given radius. This acts as an upper limit because the observations only show the tangential distance from the radio core, and not the line-of-sight distance. It is reasonable to assume the flux-to-mass ratio is constant within each galaxy, so this also serves to illustrate the fraction of the molecular mass which is contained within a given radius.}
    \label{fig:flux_integral_plot}
\end{figure*}

Since the absorbing gas in Figs. \ref{fig:main_fig_1} and \ref{fig:main_fig_2} is observed against the host galaxy's bright radio core, we can be certain of its location in terms of its RA and Dec. In four galaxies in our sample -- S555, Abell 2390, Abell 2597 and NGC 5044 -- we find the absorbing gas is likely in the galaxy centre. 

In the other four galaxies, the absorption is most likely caused by typical clouds within the galaxy. These could be at any radius at which molecular emission is present, but we now attempt to constrain the location of the absorbing gas by isolating the emission which has velocities equal to the absorbing gas.

In Fig. \ref{fig:flux_integral_plot}, we show the flux contained within increasing radii from the core as a fraction of the total flux. This is done using the two flux density maps for each of the sources shown in Figs. 1 and 2, i.e. using one map made with the emission present at all velocities, and one made with the emission present at velocities matched to the absorption. In Table \ref{tab:half-light-radii} we show half-light radii derived from Fig. \ref{fig:flux_integral_plot}, and the maps shown in Figs. \ref{fig:main_fig_1} and \ref{fig:main_fig_2}. These values serve as estimates of the median radius at which a typical molecular gas cloud will be, and a gas cloud with the same velocity as the absorbing gas. Inferences from this must be treated with caution because the curves in Fig. \ref{fig:flux_integral_plot} are found using molecular emission seen in a tangential plane to the one in which the absorbing gas lies. Assuming the gas within these galaxies is approximately spherically symmetric, the half-light radii serve as reasonable estimates for the most likely location of the absorbing gas clouds.

\begin{table}
\caption{Half-light radii in kpc for CO emission seen at (i) all velocities and (ii) velocities matched to the absorption lines seen against each galaxy's radio core. These values are found using the maps shown in Figs. \ref{fig:main_fig_1} and \ref{fig:main_fig_2}. See also Fig. \ref{fig:flux_integral_plot}.}
	\centering
	\begin{tabular}{lccccr}
\hline
 & All velocities & Absorption velocities \\
 \hline
NGC 6868 & 0.11 & 0.15 \\
S555 & 0.24 & 0.64 \\
Hydra-A & 1.2 & 0.56 \\
Abell 2390 & 7.9 & 8.9 \\
Abell 1644 & 2.4 & 6.6 \\
IC 4296 & 0.13 & 0.11 \\
Abell 2597 & 11 & 2.0 \\
NGC 5044 & 0.39 & 0.26 \\
\hline 
	\end{tabular}
    \label{tab:half-light-radii}
\end{table}

\subsection{The infall velocities of molecular clouds}

We have proposed that four systems in Figs. \ref{fig:main_fig_1} and \ref{fig:main_fig_2} are likely to contain absorption due to molecular gas clouds close to the centre of the host galaxy. Due to their high inward velocities, in time these clouds are likely to be accreted by the AGN.

However, as the moments maps show, these clouds have velocities which are similar to many others visible in emission up to several kpc out from the galaxy centre. This similarity in the velocities of two cloud populations in very different environments may be considered surprising, and some mechanism may be acting to prevent the clouds accreted by an AGN from having strongly dissimilar velocities to those which are considerably further out. Simulations suggest that this is the case, with inward radial motions significantly higher than the projected velocities of our absorption lines (i.e. intrinsic velocities of above 1000 km/s) rarely occurring at radii above 1 pc. When they do occur, these velocities are held only by the few clouds which have experienced major inelastic collisions \citep[][]{Gaspari2013}.

The clouds we infer to be accreting are detected against the galaxy's central continuum source, and have inward line-of-sight velocities of hundreds of km/s. However, they will also have tangential velocity components and will therefore not be in free-fall. Instead, they will be on more stable elliptical orbits which can decrease over time due to losses of energy and momentum \citep[e.g. by cloud-cloud collisions][]{Gaspari2013}. Due to the proximity to the galaxy centre, the outward force of ram-pressure stripping can also act to dampen any increases in velocity towards the galaxy centre caused by the strong gravitational field \citep[][]{Combes2018}. This places an upper limit on the infall velocity attainable by molecular clouds, and may have detectable effects when their absorption features are seen with different tracers, such as other molecular lines or H\small I\normalsize.

\subsection{Perturbed CO emission and AGN accretion}

From our sample of eight galaxies with both molecular emission and absorption, four have absorption which is likely due to clouds accreting onto their AGN (S555, Abell 2390, Abell 2597, NGC 5044). The other four have absorption which is likely due to typical clouds at much larger radii (NGC6868, Hydra-A, Abell 1644 and IC 4296). Interestingly, those with absorption likely due clouds accreting onto the AGN have irregular velocity structures (Fig. \ref{fig:velocity_and_dispersion_maps}), while those with absorption due to typical clouds have smooth molecular emission profiles. 

An irregular and disturbed velocity structure in a galaxy's molecular emission may therefore be conducive to fuelling of the AGN. This would likely be due to increased cloud-cloud collisions, which cause a loss of angular momentum and lead to clouds falling to lower orbits. 

In the three galaxies in our sample which contain emission but no absorption (Fig. \ref{fig:maps_of_emitters}), one has a smooth velocity structure (RXC J1350.3+0940), while two are irregular (MACS 1931.8-2634 and RXC J1603.6+1553). Although there is no detectable absorption in these systems, we have provided upper limits in Table \ref{tab:gas_densities}. By comparing these upper limits with the absorption column densities in the other galaxies, we can judge whether or not their lack of absorption is due to the sensitivity of the observations. 

Where we find absorption due to typical clouds, the column densities (1100, 210, 310, 290 M$_{\odot}$/pc$^{2}$) are lower than in cases where the absorption is due to clouds undergoing AGN accretion (1300, 15000, 6800, 3800 M$_{\odot}$/pc$^{2}$). For the systems in which there is molecular emission only, the absorption upper limits are (<700, 2200, and <3400 M$_{\odot}$/pc$^{2}$). Assuming the dichotomy in the absorption column densities of the two types of absorber is more than coincidence, these upper limits mean we cannot rule out the presence of undetected absorption due to typical clouds. However, there is a much lower probability that they contain absorption due to clouds accreting onto their AGN. Two of these three systems contain a disturbed velocity structure, so deep and wide absorption due to accreting clouds is unlikely to be a universal feature of galaxies with irregular velocity structures.

\subsection{Limitations to our classifications}

So far we have restricted our classification of these absorption systems, suggesting they are likely due to alignments between the continuum and (i) a small fraction of the clouds we see in emission, or (ii) clouds near the galaxy centre which are in the process of accreting onto the AGN. However, there is no reason to believe a dichotomy exists i.e. that no systems combine these two types of absorption. The X-ray selected \citet{Rose2019b} survey in which most of the absorption systems we present were identified had a detection rate of around 40 per cent. Given that we find a roughly equal prevalence of the two types of absorber, the detection rate of each can be approximated to 20 per cent. If they can be treated as independent, absorption of both types will be detectable in around four per cent of sources (though this could rise with more sensitive observations).

In Fig. \ref{fig:detection_factors} we noted differences between the clumping factors of the two types of absorber. Systems with absorption due to accreting clouds have much higher clumping factors than those due to the galaxy's molecular clouds which are visible in emission (Fig. \ref{fig:detection_factors}). If a system's absorption is due to typical clouds within the galaxy, high clumping factors may also be caused by significant inhomogeneities in the ISM. Therefore, we do not yet have reason to believe that high clumping factors are an integral or unique property of systems which have molecular absorption due to clouds accreting onto their AGN. With greater sensitivity, we may find absorption with high velocity dispersions and high velocities towards the galaxy centre, but which are relatively weak.

\section{Conclusions}

We have analysed ALMA observations of eight galaxies with both molecular emission and absorption and constrained the location of their absorbing gas. We also show observations of three massive galaxies with only molecular emission lines. Our findings can be summarized as follows: 
\begin{itemize}
    \item We introduce the `clumping factor' of molecular absorption line systems -- the ratio of two column densities along the line-of-sight to a galaxy's continuum source. The first is estimated from the molecular emission spatially coincident with the radio core. This gives the amount of gas from across the galaxy which covers a typical line of sight to the continuum source. The second is the actual column density of molecular gas covering the continuum, estimated from the absorption lines.
    \item When comparing the clumping factors and the FWHM of the absorption lines in our sample with the velocities of their molecular absorption lines (Fig. \ref{fig:detection_factors}), two distinct populations become apparent. One has low clumping factors, lower FWHM and molecular absorption at close to 0\,km/s. The others have high clumping factors of $\gtrsim$10, higher FWHM and the absorbing gas has high inward velocities towards the galaxy centre.
    \item We argue that those with low clumping factors, lower FWHM and absorption close to 0\,km/s are due to chance alignments between the continuum source and a fraction of the same molecular clouds responsible for the galaxy's molecular emission. The low velocity dispersions in these systems are similar to those found for absorption far out from galaxy centres using background quasars. Additionally, since the absorption is due to chance alignments between the continuum source and a fraction of the molecular clouds which are visible in emission, the absorption is always embedded within that emission.
    \item Absorbers with high clumping factors, higher FWHM and high inward velocities are likely created by gas clouds accreting onto their host AGN. The absorption in these systems is unlikely to have been produced in the same manner as described above for two main reasons. Firstly, the clumping factors consistently imply that at least an order of magnitude more absorption is being detected than is predicted from the emission. Secondly, the absorption lines are significantly redshifted compared with the molecular emission (except in Abell 2390, which has 900 km/s wide emission). If the absorption was produced by the same population of clouds responsible for the emission, absorption would be most likely to occur at or near the peak of that emission. 
    \item Clouds accreting onto their AGN are much less numerous than those throughout a galaxy which make up the molecular emission profile. Their scarcity means they are not visible in emission, while their proximity to the galaxy centre means they are relatively more likely to align with the continuum source.
    \item We present the first molecular mass estimates for five brightest cluster galaxies. Three of these have molecular absorption lines against their central continuum source (NGC6868, S555, A2390) and two do not (RXC J1350.3+0940, RXC J1603.6+1553). 
    \item Three of the brightest cluster galaxies we present have molecular masses of $>10^{10}\textnormal{M}_{\odot}$, but lack molecular absorption lines. This is surprising given their masses are two or more orders of magnitude higher than in several other absorbing sources. The column density of molecular gas covering the continuum source is a better predictor of absorption due to alignments between the continuum and a fraction of the clouds visible in emission.
\end{itemize}

\section*{Acknowledgements}
T.R. thanks the Waterloo Centre for Astrophysics and generous funding to B.R.M. from the Canadian Space Agency and the National Science and Engineering Research Council of Canada. A.C.E. acknowledges support from STFC grant ST/P00541/1. M.G. acknowledges partial support by HST GO-15890.020/023-A and the \textit{BlackHoleWeather} program. HRR acknowledges support from an STFC Ernest Rutherford Fellowship and an Anne McLaren Fellowship. P.S. acknowledges support by the ANR grant LYRICS (ANR-16-CE31-0011).

This paper makes use of the following ALMA data: 2017.1.00629.S, 2021.1.00766.S, 2016.1.01214.S, 2012.1.00988.S, 2018.1.01471.S, 2011.0.00735.S and 2015.1.01572.S. ALMA is a partnership of ESO (representing its member states), NSF (USA) and NINS (Japan), together with NRC (Canada), MOST and ASIAA (Taiwan), and KASI (Republic of Korea), in cooperation with the Republic of Chile. The Joint ALMA Observatory is operated by ESO, AUI/NRAO and NAOJ. In addition, publications from NA authors must include the standard NRAO acknowledgement: The National Radio Astronomy Observatory is a facility of the National Science Foundation operated under cooperative agreement by Associated Universities, Inc.

This research made use of \texttt{Astropy} \citep{the_astropy_collaboration_astropy_2013,the_astropy_collaboration_astropy_2018}, \texttt{Matplotlib} \citep{hunter_matplotlib_2007}, \texttt{numpy} \citep{walt_numpy_2011,harris_array_2020}, \texttt{Python} \citep{van_rossum_python_2009}, \texttt{Scipy} \citep{jones_scipy_2011,virtanen_scipy_2020} and \texttt{Aplpy} \citep[][]{aplpy}. We thank their developers for maintaining them and making them freely available.

%%%%%%%%%%%%%%%%%%%%%%%%%%%%%%%%%%%%%%%%%%%%%%%%%%
\section*{Data Availability}

The vast majority of the data underlying this article are publicly available from the ALMA archive. The CO(2-1) observation of NGC6868 will become publicly available on 2022 November 25. Before then, it will be shared on reasonable request to the corresponding author.

%%%%%%%%%%%%%%%%%%%%%%%%%%%%%%%%%%%%%%%%%%%%%%%%%%

%%%%%%%%%%%%%%%%%%%% REFERENCES %%%%%%%%%%%%%%%%%%

% The best way to enter references is to use BibTeX:

\bibliographystyle{mnras}
%\bibliography{bibliography} % if your bibtex file is called example.bib

%%%%%%%%%%%%%%%%%%%%%%%%%%%%%%%%%%%%%%%%%%%%%%%%%%

%%%%%%%%%%%%%%%%% APPENDICES %%%%%%%%%%%%%%%%%%%%%

\appendix

%%%%%%%%%%%%%%%%%%%%%%%%%%%%%%%%%%%%%%%%%%%%%%%%%%

% Don't change these lines
\bsp	% typesetting comment
\label{lastpage}
\end{document}